\documentclass[%
reprint,
groupedaddress,
nofootinbib,
 amsmath,amssymb,
 aps,
prb,
floatfix,
]{revtex4-2}

\usepackage{graphicx}	% Include figure files
\usepackage{bm}	% bold math
\usepackage{amssymb}
\usepackage{multirow}
\usepackage{array}
\usepackage{color,soul}
\usepackage{float}
\usepackage{tabularx}
\usepackage{xr}

\begin{document}
\title{Nature of Temperature Chaos in Spin Glasses}
\author{Hongze~Li}
\author{Jiaming~He}
\author{Raymond~L.~Orbach}
\email[Contact author: ]{orbach@utexas.edu}

\affiliation{Texas Materials Institute, The University of Texas at Austin, Austin, Texas  78712, USA}
\date{\today}   % It is always \today, today}

\begin{abstract}
\noindent Temperature chaos (TC) in spin glasses has been claimed to exist no matter how small the temperature change, $\Delta T$.  However, experimental studies have exhibited a finite value of $\Delta T$ for a transition to TC. This paper explores the onset of TC with much higher resolution than before and over a larger temperature range.  We find that TC is always present, though small at the smallest $\Delta T$ that we can reliably measure.  However, it grows rapidly as $\Delta T$ increases, the region of rapid growth coinciding with the $\Delta T$ predicted from renormalization group arguments and observed experimentally.  We are able to transcend the full range of TC, from the completely reversible state to one that is maximally decorrelated from the initially prepared state.
\end{abstract}

%\keywords{Suggested keywords}	%Use showkeys class option if keyword
\maketitle

\maketitle
\section{\label{sec:introduction}Introduction}
Temperature chaos (TC) in spin glasses has been controversial ever since it was first introduced \cite{McKay1982,Bray:87,Fisher:86,Fisher:88}. The simplest description of TC is ``...the complete reorganization of the equilibrium configurations by the slightest change in temperature''  \cite{Fernandez:2013}.  Concomitantly, extending the analysis to off-equilibrium dynamics, TC ``...means that the spin configurations that are typical from the Boltzmann weight at temperature $T_1$ [we refer to this temperature as $T_{\text {initial}}$] are very atypical at temperature $T_2$ [we refer to this temperature as $T_{\text {final}}$] no matter how close the two temperatures $T_1$ and $T_2$ are'' \cite{Baity:21}.  A rather complete list of theoretical papers supporting or questioning the existence of TC can be found in \cite{zhai:22}.  Experimentally, the three experimental papers that have reported TC in spin glasses, \cite{PhysRevLett.89.097201, Guchhait:15, zhai:22}, observe the onset of TC only at a relatively large temperature change, close to that predicted from renormalization group arguments \cite{McKay1982,Bray:87}.

This paper probes the presence of TC at a higher resolution level and exhibits TC for even the smallest $\Delta T=T_{\text {initial}}-T_{\text {final}}$ that we can reliably measure.  We find its magnitude to be small at the smallest temperature change, $\Delta T$, resolving the apparent contradiction between theory and experiment. TC does become large at a crossover temperature close to that predicted by the arguments of the renormalization group \cite{McKay1982,Bray:87}, and in the range of $\Delta T$ reported in previous experimental reports \cite{PhysRevLett.89.097201, Guchhait:15, zhai:22}.  We find TC to continue to increase with increasing $\Delta T$ until the entirety of the spin glass system is maximally decorrelated from the initially prepared state.

Previous work \cite{Freedberg:23,Paga:23} has quantitatively described the nature of TC in spin glasses.  Their analysis begins with the preparation of the spin glass at some initial temperature, $T_{\text {initial}}$.  The spin glass is aged for a time $t_{{\text {w}},{\text {initial}}}$ allowing the growth of the correlation length from nucleation to $\xi(t_{{\text {w}},{\text {initial}}},T_{\text {initial}})$.  The temperature is then dropped to $T_{\text {final}}$ and the spin glass is aged for $t_{{\text {w}},{\text {final}}}$, allowing a new correlation length to grow from nucleation to $\xi(t_{{\text {w}},{\text {final}}},T_{\text {final}})$.  Because of the rapid slowing down of the growth of the spin glass correlation length with decreasing temperature \cite{Sibani:93,Kisker:96,Marinari:96}, the correlations created at $T_{{\text {initial}}}$ are essentially frozen at $T_{\text {final}}$.  The associated correlated volume, subtended by $\xi(t_{{\text {w}},{\text {initial}}},T_{\text {initial}})$, interferes with the growth of $\xi(t_{{\text {w}},{\text {final}}},T_{\text {final}})$, leading to quantitative predictions for rejuvenation and the magnitude of memory. The reader is referred to \cite{Freedberg:23,Paga:23} for full details.  The basic assumption underlying their analysis, and our own in the following, is that TC results from an {\it independent} growth of a chaotic regime at $T_{\text {final}}$.

The protocol utilized in this paper generates a direct measure of the amount of TC created upon a reduction of temperature from $T_{\text {initial}}$ to $T_{\text {final}}$.  The renormalization group analysis of Bray and Moore \cite{Bray:87} defines a chaos length,
\begin{equation}
\ell_c(T_{\text {initial}}-T_{\text {final}})\approx a_0\, \bigg[ {\frac {T_{\text {initial}}}{T_{\text {initial}}-T_{\text {final}}}}\bigg]^{1/\zeta},
\end {equation}
where $a_0$ is a characteristic lattice dimension (e.g. the mean distance between Mn ions in a CuMn dilute alloy), $\zeta$ is the chaos exponent $\zeta=d_s/2-\theta$, $d_s$ the surface fractal dimension and $\theta$ the replicon exponent.  Typically, $\zeta$ is found to be approximately unity \cite{zhai:22}.
The condition for TC from \cite{Bray:87} is,
\begin {equation}
\ell_c(T_{\text {initial}}-T_{\text {final}})=\xi(t_{{\text {w}},{\text {initial}}},T_{\text {initial}}).
\end{equation}
This condition has been verified in some detail for a 6 at.\% CuMn single crystal in \cite{zhai:22}.  However, this condition is a crossover, and TC exists on either side of the temperature drop exhibited in Eq. (2).  An example can be found in Fig. 3 of Baity-Jesi et al. \cite {Baity:21}.

Temperature chaos implies that the state at $T_{\text {final}}$ has no relationship to the state created at 
$T_{\text {initial}}$.  However, Hammann et al. \cite{Hammann:92} showed that for $T_{\text {initial}}-T_{\text {final}}$ sufficiently small, this was not true. On the one hand, they demonstrated that for temperature differences $\leq$ 60 mK, they could reproduce the state initially prepared at $T_{\text {initial}}$ by cycling the temperature from $T_{\text {initial}}$ to $T_{\text {final}}$ and then back to $T_{\text {initial}}$. This represents reversible behavior upon a reduction in temperature, and is inconsistent with the creation of a chaotic state.  On the other hand, they were unable to reproduce the state initially prepared at $T_{\text {initial}}$ for temperature differences $\geq$ 80 mK. At the time no explanation was given for this difference in behavior.  It is the thesis of this paper that this was probable evidence for the onset of TC. 

Our procedure to examine the onset and behavior of TC as a function of the difference in temperature is the following.  First, the spin glass is cooled from above $T_\text{g}$ (here, $T_\text{g} \approx$ 31.5 K) to an initial temperature $T_{\text {initial}}$ and aged for a time $t_{{\text {w}},{\text {initial}}}$.  This sets a length scale in that the spin glass correlation length will have grown to $\xi(t_{{\text {w}},{\text {initial}}},T_{\text {initial}})$. Upon lowering the temperature to $T_{\text {final}}$ and aging for the time $t_{{\text {w}},{\text {final}}}$, two changes occur.  The first is a reversible behavior seen by Hammann et al. \cite{Hammann:92}.  It will turn out to be a rather complex quantity to calculate and is described in some detail below.  

The second follows from the findings of the simulations of Baity-Jesi et al. \cite{Baity:21}.  They showed that upon a change in temperature, a probability distribution for TC (they termed it the ``chaotic parameter'' following \cite{Ney-Nifle:97}) was present that echoed the length-scale formulation of Bray and Moore \cite{Bray:87}, Eq. (1).  Thus, upon a change in temperature, $\Delta T$, there is a probability that a portion of the spin glass is maximally decorrelated from that created at $T_{\text {initial}}$.  The probability depends upon the proximity of $\ell_c(T_{\text {initial}}-T_{\text {final}})$ to $\xi(t_{{\text {w}},{\text {initial}}},T_{\text {initial}})$.  If, on the one hand, $\xi(t_{{\text {w}},{\text {initial}}},T_{\text {initial}})\ll$$\ell_c(T_{\text {initial}}-T_{\text {final}})$, as would be the case for small $\Delta T$, the probability of TC would depend upon the ``wings'' of the chaotic parameter (remember that Eq. (2) is a crossover condition) and be small.  If, on the other hand, $\xi(t_{{\text {w}},{\text {initial}}},T_{\text {initial}})\sim$ $\ell_c(T_{\text {initial}}-T_{\text {final}})$, the chaotic parameter would be close to unity, and the probability for TC would be large.

In summary, upon a temperature change $\Delta T$, {\it both} reversible and chaotic behavior are present.  The relative magnitudes of the two will be extracted through the experimental protocol described below.

The method we use relies on extracting the characteristic response time defined as the time when the relaxation function,
\begin{equation}
S(t) = dM(t,t_{\text {w}};H)/d{\text {ln}}t,  
\label{eq:s_of_t}
\end{equation}
peaks from time dependent magnetization measurements.  This time, denoted at $t_{\text {w}}^{\text {eff}}$, is close to the impressed waiting time $t_{\text {w}}$, and has been the standard means of extracting the response time for spin glass dynamics ever since it was introduced by Nordblad et al. \cite{Nordblad:86}. 

Some representative samples of $S(t)$ for our single crystal CuMn 6 at.\% sample can be found in Appendix A, Figs. 7 and 9.  The width of the $S(t)$ curve as a function of time $t$ makes it difficult to extract a precise value for the time at which $S(t)$ peaks.  Our method for achieving the necessary accuracy is described in Appendix A. We also check magnetic field linearity in Appendix B.

The extracted $t_{\text {w}}$ produces an important physical quantity. The magnitude of the largest free energy barrier, $\Delta_{\text {max}}(t_{\text {w}},T)$ , generated by the growth of the spin glass correlation length, is connected to $t_{\text {w}}^{\text {eff}}$ through the Arrhenius relation,
\begin{equation}
t_{\text {w}}^{\text {eff}}=\tau_0\,{\text {exp}}[\Delta_{\text {max}}(t_{\text {w}},T)/k_BT],    
\end{equation}
where $\tau_0$ is a typical exchange time, usually taken as $\tau_0=\hbar/k_BT_{\text {g}}$. This relationship is based on a hierarchical organization of metastable states, first articulated by Refregier et al. \cite{Refregier:87}.  A more recent publication titled ``Real Spin Glasses Relax Slowly in the Shade of Hierarchical Trees'' \cite{Vincent:09} is particularly convincing in this respect.

In order to separate the amplitudes of the chaotic and reversible states, it is necessary to carry out the following two respective experimental protocols.

\subsubsection{Amplitude of the chaotic component}
We cool the spin glass from above $T_{\text{g}}$, the spin glass transition temperature, to an initial temperature $T_{\text {initial}}$ and age the system for a time $t_{{\text {w}},{\text {initial}}}$.  We lower the temperature (as rapidly as possible) to $T_{\text {initial}}$ and age again for $t_{{\text {w}},{\text {final}}}$.  We apply a magnetic field and measure the time dependence of the (slowly) increasing magnetization, from which we extract an effective response time $t_{{\text {w}},T_{\text {initial}}\rightarrow T_{\text {final}}}^{\text {eff}}$ (see below for details; the values are listed in the third column of Table I). This time contains the contribution of both the chaotic and reversible components for the smallest reductions in temperature.  The latter will need to be subtracted from the former to extract the amplitude of the chaotic state in this temperature range.  However, as will be shown below, the effective response time for the reversible component will become very large as the reduction in temperature increases, leaving behind the time at which $S(t)$ peaks to be that from the chaotic component. This is consistent with the notion that TC means that the state at $T_{\text {final}}$  has no relationship to the state created at $T_{\text {initial}}$.

If this is true, then $t_{{\text {w}},T_{\text {initial}}\rightarrow T_{\text {final}}}^{\text {eff}}$ should approach the time at which $S(t)$ peaks for a direct reduction of temperature from $T_{\text {initial}}$ to $T_{\text {final}}$ after aging for $t_{{\text {w}},{\text {final}}}$. To be consistent with the literature \cite{Paga:23}, we term this effective response time $t_{{\text {w}},{\text {native}}}^{\text {eff}}$, which is listed in column 1 of Table I. A glance at the two columns in Table I should convince the reader that indeed we have entered the chaotic state for sufficiently large reductions in temperature (the slight difference will be discussed in detail below).

\subsubsection{Amplitude of the reversible component}
At first sight, this should be an easy value to calculate, given Eq. (4).  One would simply compute $t_{\text {w}}^{\text {eff}}$ by changing the temperature from $T_{\text {initial}}$ to $T_{\text {final}}$.  Unfortunately, this is not true because, as Hammann et al. \cite{Hammann:92} have shown, $\Delta_{\text {max}}$ itself is temperature dependent.  Their finding is exhibited below in Fig. 1:
\begin{figure}[htbp]
    \centering
    \includegraphics[width=0.4\textwidth]{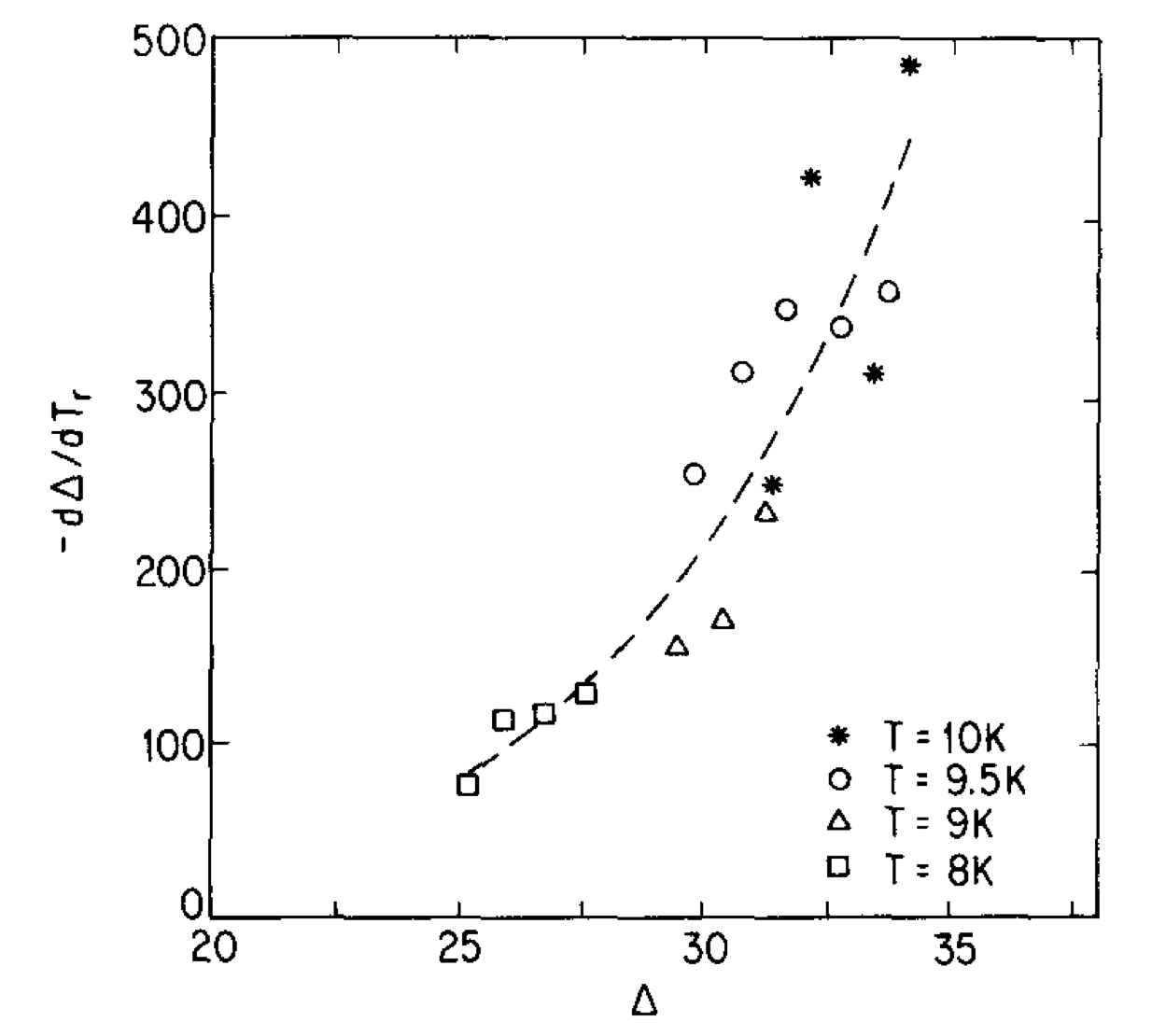}
    \caption{A plot of $d\Delta/dT_{\text {r}}$ vs $\Delta$, where $T_{\text {r}}=T/T_{\text {g}}$, for each of four values of $T$, for a Ag:Mn 2.6 at.\% sample. The set of data points define a unique curve [the dashed line, Eq. (3)], the points at each fixed value of $T$ are consistent with this unique curve, and the curve is therefore independent of temperature.  Reproduced from Fig. 4 of Ref. \cite{Hammann:92}}
    \label{fig:Hammann}
\end{figure}
The remarkable feature of Fig. 1 is that the data are independent of temperature, and can be fitted, with equally acceptable results, to a power law,
\begin{equation}
-d\Delta/dT_{\text {r}}=a\Delta^n,
\label{eq:d_Delta_d_T}
\end{equation}
with $a = 2.9\times 10^{-7}$ and $n = 6$, as shown by the dashed line in Fig. \ref{fig:Hammann}, or to an exponential form,
\begin{equation}
-d\Delta/dT_{\text {r}}=\alpha \text{exp}(\beta\Delta),
\label{eq:d_Delta_d_T_exp}
\end{equation}
with $\alpha = 0.5$ and $\beta = 0.2$ \cite{Hammann:92}.

It is important to analyze the import of Fig. 1. The relationships, Eq. (3) and (4), are a result of the {\it independence} of $-d\Delta/dT_r$ with respect to the measurement temperature.  This is a direct result of the self-similar structure of the metastable states governing spin glass dynamics \cite{Hammann:92}.  That is, as the measurement temperature is changed, the accessible free energy state structure remains the same.  Hence, knowing $\Delta_{\text {max}}(t_{\text {w}},T)$ at any temperature (say, $T_{\text {initial}}$) enables the calculation of $\Delta_{\text {max}}(t_{\text {w}},T)$ at any other temperature (say $T_{\text {final}}$).  Our protocol is to measure $\Delta_{\text {max}}(t_{\text {w}}=10^4~{\text {sec}},T=T_{\text {initial}})$ by finding the time at which the respective $S(t)$ peaks through Eq. (4). We call this time $t_{{\text {w}},r}^{\text {eff}}$. We then find $\Delta_{\text {max}}(t_{\text {w}}=10^4\,{\text {sec}},T_{\text {initial}})$ through Eq. (4).  In summary, we have determined, at $T_{\text {initial}}$, $\Delta_{\text {max}}(t_{\text {w}}=10^4\,{\text {sec}},T_{\text {initial}})$, which was generated as a consequence of a measured effective response time of $t_{{\text {w}},r}^{\text {eff}}$, and then we use Eq. (5) to find the magnitude of {\it that} barrier at the lower temperature of $T_{\text {final}}$.

However, we wish to put the reversible effective response time on the same footing as the effective response time for the chaotic component. That means we need to age from that barrier at the lower temperature of $T_{\text {final}}$ for the aging time $t_{\text {w}}=10^4\,{\text {sec}}$ and then add the total effective aging times at $T_{\text {initial}}$ and $T_{\text {final}}$.  We have done the former (i.e. $t_{{\text {w}},r}^{\text {eff}}$), but how do we find the latter?  It is simply what we have already calculated as $t_{{\text {w}},{\text {native}}}^{\text {eff}}$.  That quantity was the result of a direct reduction of temperature from above $T_{\text {g}}$ to $T_{\text {final}}$, so that the nucleation began precisely at the maximum barrier height associated with $T_{\text {final}}$.  Therefore, all we need to do to calculate the effective response time for the reversible component at $T_{\text {final}}$ is to sum $t_{{\text {w}},r}^{\text {eff}}$ and $t_{{\text {w}},{\text {native}}}^{\text {eff}}$, which we refer to as $t_{{\text {w}},R}^{\text {eff}}$ and enter into column 4 of Table I.

In sum, we have outlined the protocol for extracting the contributions of the chaotic and reversible components of the final state at $T_{\text {final}}$.  The time for which $S(t)$ peaks is measured and reported in column 3 of Table 1, and the time at which the $S(t)$ peaks for the reversible component is reported in column 4 of Table 1.  Our analysis of the significance of these peak times is contained in the next section, which reports the actual measurements and analysis.

\section{\label{sec:analysis}Measurements and Analysis}
\begin{table*}[htbp]
    \centering
    \caption{Listing of measured and calculated values for the temperature $T_{\text {final}}$ (column 1); the measured native effective response time at $T_{\text {final}}$ for $t_{{\text {w}},{\text {initial}}}=10^4$ s (column 2); the measured effective response time after $t_{{\text {w}},{\text {initial}}}=10^4$ s at $T_{\text {initial}}$, dropping the temperature to $T_{\text {final}}$, and waiting for $t_{{\text {w}},{\text {final}}}=10^4$ s (column 3); the calculated effective response time of the reversible (cumulative) portion (column 4); and the calculated native correlation length with $t_{{\text {w}},{\text {initial}}}=10^4$ s (column 5).  Our calculations use the power law form, Eq. (\ref{eq:d_Delta_d_T}), for the temperature dependence of $\Delta$.}
    \begin{ruledtabular}
        \begin{tabular}{cccccccc}
       $T$(K) &$t_{{\text {w}},{\text {native}}}^{\text {eff}}$($\times~10^4$~s) & $t_{{\text {w}},T_{\text {initial}}\rightarrow T_{\text {final}}}^{\text {eff}}$($\times~10^4$ s) & $t_{\text {\text {R}}}^{\text {eff}}$(s) & $\xi/a_0$\\
        \hline
        18.00  & 1.84 $\pm$ 0.04 & -  & - & 9.715\\
        17.95 & 1.85 $\pm$ 0.02 & 3.92 $\pm$ 0.01 & 4.10~$\times~10^4$ & 9.653\\
        17.90  & 1.90 $\pm$ 0.02 & 4.38 $\pm$ 0.06 & 4.67~$\times~10^4$ & 9.593\\
        17.80  & 1.90 $\pm$ 0.03 & 5.42 $\pm$ 0.10 & 6.10~$\times~10^4$ & 9.472\\ 
        17.70  & 1.91 $\pm$ 0.02 & 6.16 $\pm$ 0.02 & 8.31~$\times~10^4$ & 9.353\\
        17.55  & 1.81 $\pm$ 0.01 & 9.21 $\pm$ 0.06 & 1.40~$\times~10^5$ & 9.178\\
        17.50  & 1.89 $\pm$ 0.05 & 10.04 $\pm$ 0.03 & 1.70~$\times~10^5$ & 9.120\\
        17.30  & 1.87 $\pm$ 0.02 & 13.23 $\pm$ 0.05 & 3.81~$\times~10^5$ & 8.708\\
        17.00  & 1.84 $\pm$ 0.01 & 10.68 $\pm$ 0.36 & 1.42~$\times~10^6$ & 8.387\\
        16.50  & 2.00 $\pm$ 0.06 & 5.08 $\pm$ 0.04 & 1.49~$\times~10^7$ & 7.722\\
        16.00  & 1.92 $\pm$ 0.01 & 2.78 $\pm$ 0.05 & 1.83~$\times~10^8$ & 7.258\\
        15.50  & 1.90 $\pm$ 0.03 & 1.90 $\pm$ 0.03 & 2.66~$\times~10^9$ & 6.695\\
        15.00  & 1.89 $\pm$ 0.01 & 1.73 $\pm$ 0.02 & 4.60~$\times~10^{10}$ & 6.183\\
        14.50  & 1.94 $\pm$ 0.03 & 1.66 $\pm$ 0.01 & 9.70~$\times~10^{11}$ & 5.717\\
        14.00  & 2.09 $\pm$ 0.01 & 1.77 $\pm$ 0.02 & 2.54~$\times~10^{13}$ & 5.384\\
        13.00  & 1.84 $\pm$ 0.07 & 1.53 $\pm$ 0.02 & 3.71~$\times~10^{16}$ & 4.625\\
        12.00  & 1.87 $\pm$ 0.03 & 1.59 $\pm$ 0.03 & 1.82~$\times~10^{20}$ & 4.052\\
        11.00  & 1.82 $\pm$ 0.01 & 1.51 $\pm$ 0.05 & 4.20~$\times~10^{24}$ & 3.558\\
        6.00 & 1.62 $\pm$ 0.04 & 1.30 $\pm$ 0.01 & 1.78~$\times~10^{68}$ & 1.913\\
        \end{tabular}
    \end{ruledtabular}  
    \label{tab:results}
\end{table*}

The spin glass CuMn single crystal sample (6 at.\% Mn, $T_{\text {g}} = 31.5$ K) is cooled from above $T_{\text {g}}$ to $T_{\text {initial}} =18.00$ K, a magnetic field of $H=100$ Oe is applied after an aging time of $10^4$ s, and the magnetization change with time recorded. $S(t)$ is created from Eq. (1), and the characteristic response time $t_{{\text {w}},{\text {native}}}^{\text {eff}}$ is extracted from the time at which $S(t)$ peaks.  It is listed in the first row, second column, of Table \ref{tab:results}.  Its value, using the Arrhenius relation of Eq. (2), generates $\Delta_{\text {max}}= 22.21~T_{\text {g}}$. We then follow the same procedure for a series of temperatures $T_{\text {final}}$, from $T_{\text {final}}=17.95$ K to as low at $T_{\text {final}}=6.00$ K.  The values for $t_{\text {w,native}}^{\text {eff}}$ for each $T_{\text {final}}$ are listed in column 2 of Table \ref{tab:results}.

The spin glass single crystal sample is now cooled from above $T_{\text {g}}$ to $T_{\text {initial}} = 18.00 $ K.  The system is aged at $T_{\text {initial}}$ for $10^4$ s, the temperature is dropped to $T_{\text {final}}$, and the system is aged for another $10^4$ s. After the second aging, a magnetic field of 100 Oe is applied, and the magnetization change is recorded as a function of time. The $S(t)$ curves are generated through Eq. (1), and the effective response times, $t_{t_{\text {w}},T_{\text {initial}}\rightarrow T_{\text {final}}}^{\text {eff}}$ for each $T_{\text {final}}$ are extracted from the time at which the respective $S(t)$ peaks.  Their values are listed in column 3 of Table \ref{tab:results}.  If the characteristic response time for the reversible component of the state at $T_{\text {final}}$ is well beyond the time scale of the experiment (e.g. for $T_{\text {final}}\leq 16$ K), the remaining state is maximally decorrelated from that state prepared at $T_{\text {initial}}$.  It is chaotic with respect to the state at $T_{\text {initial}}$.

\begin{figure}[htbp]
    \centering
    \includegraphics[width=0.48\textwidth]{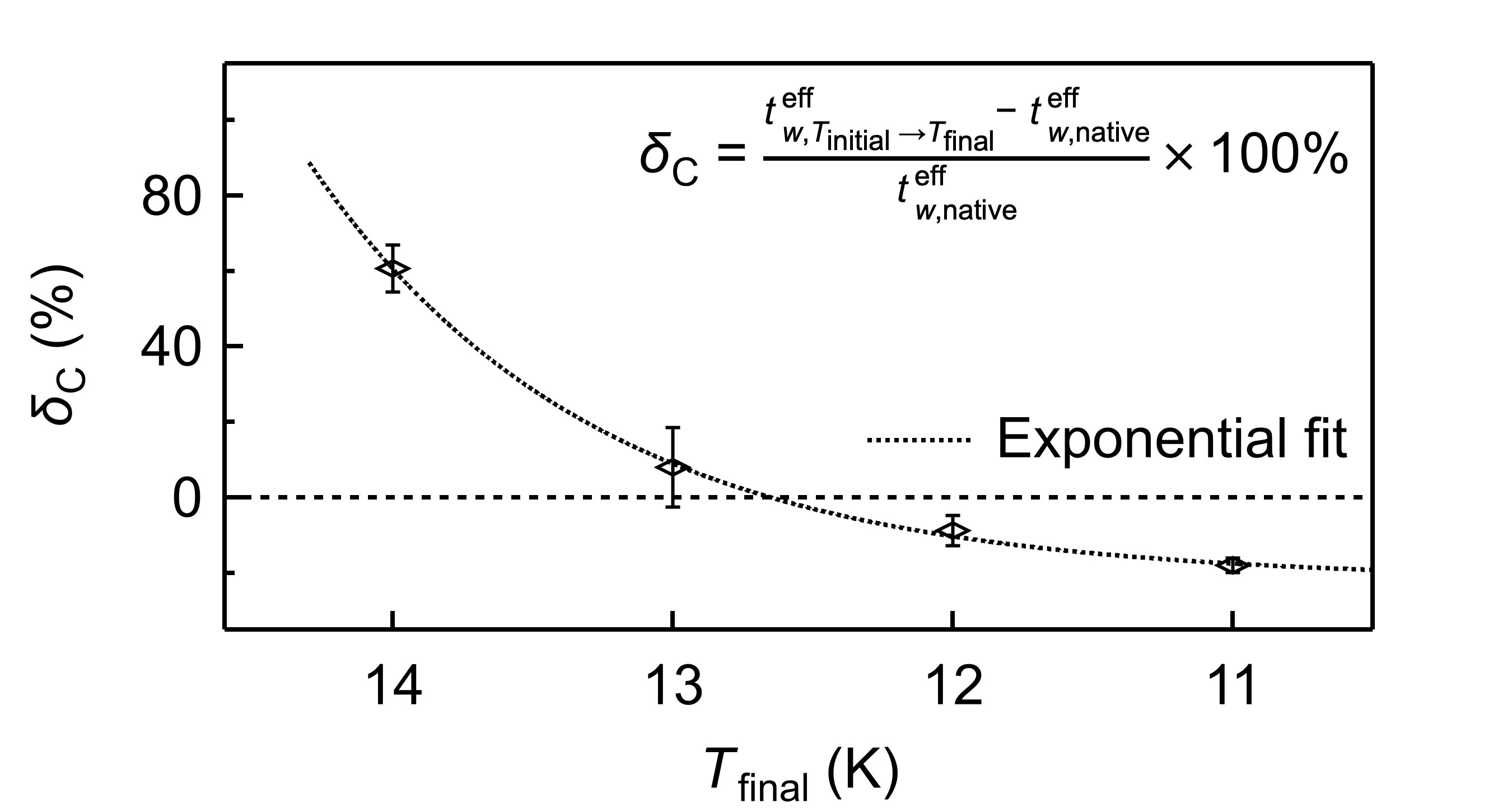}
    \caption{(a) The measured native effective response time at $T_{\text {final}}$ for $t_{{\text {w}},{\text {initial}}}=10^4$ s (diamonds) and the measured effective response time after the $T_{\text {initial}} = 18.00$ K $\rightarrow T_{\text {final}}$ protocol (circles); (b) the difference $\delta_{\text {TC}}=(t_{\text {R}}^{\text {eff}}-t_{{\text {w}},T_{\text {initial}}\rightarrow T_{\text {final}}}^{\text {eff}})/t_{{\text {w}},T_{\text {initial}}\rightarrow T_{\text {final}}}^{\text {eff}}\times 100\%$ at small $\Delta T$s ($\Delta T = T_{\text {initial}} - T_{\text {final}}$); (c) the difference $\delta_{{\mathcal {C}}}=(t_{{\text {w}},T_{\text {initial}}\rightarrow T_{\text {final}}}^{\text {eff}}-t_{{\text {w}},\text {native}}^{\text {eff}})/t_{{\text {w}},\text {native}}^{\text {eff}}\times 100\%$ at large $\Delta T$s, with the exponential fit superposed.}
    \label{fig:results}
\end{figure}

If the system is maximally decorrelated with respect to the initial state at a given $T_{\text {final}}$, the values $t_{{\text {w}},{\text {native}}}^{\text {eff}}$ (column 2) should be equal to $t_{{\text {w}},T_{\text {initial}}\rightarrow T_{\text {final}}}^{\text {eff}}$ (column 3).  Examination of Table \ref{tab:results} shows that this occurs for $T_{\text {final}}=15.50$ or $\Delta T=2.50$ K.  The latter is slightly smaller than the former for all larger $\Delta T$ for reasons that will be discussed below.  The results of these two procedures are plotted against $T_{\text {final}}$ in Fig. \ref{fig:results}(a).

The next step is to disentangle the contribution to the measured effective response times reported in column 3 of Table \ref{tab:results} from the reversible and chaotic components.  The native response time at $T_{\text {initial}}=18.00$ K is the starting point, as no temperature change has occurred, hence no TC.  The maximum barrier height, $\Delta_{\text {max}}$, is generated by the growth of the correlation length during the aging time $t_{\text {w}}=10^4$ s at 18.00 K.  Na{\"i}vely, one might choose to use the Arrhenius relation to calculate the reversible (cumulative) effective response time for each $T_{\text {final}}$.  However, that would be incorrect.  Hammann et al. \cite{Hammann:92} showed that $\Delta_{\text {max}}$ {\it increases} at the temperature is reduced. They evaluated $d\Delta_{\text {max}}/dT_{\text {r}}$ ($T_{\text {r}} = T/T_{\text {g}}$) as a function of temperature $T$.  We reproduced their findings in Fig. \ref{fig:Hammann} (their $T\equiv T_{\text {initial}}$).  We use Fig. \ref{fig:Hammann} to calculate $\Delta_{\text {max}}(t_{\text {w}},T_{\text {initial}}\rightarrow T_{\text {final}})$ for our CuMn 6 at.\% spin glass sample.  Though Fig. \ref{fig:Hammann} was for a AgMn sample, their use of reduced units suggests that we can apply it to our CuMn sample.

We attribute the difference between the value of the reversible effective response time, $t_R^{\text {eff}}$, column 4 of Table \ref{tab:results}, and the experimental value of the cycled effective response time, $t_{{\text {w}},{T_{\text {initial}}\rightarrow T_{\text {final}}}}^{\text {eff}}$, column 3 of Table \ref{tab:results}, to the presence of TC.  This is because the spin orientations initially created at $T_{\text {initial}}$ after aging for $t_{w,{\text {initial}}}$ have no relationship to those created when TC occurs.  Hence, $t_{{\text {w}},{T_{\text {initial}}\rightarrow T_{\text {final}}}}^{\text {eff}}$ represents the growth of $\xi(t_{{\text {w}},{\text {final}}},T_{\text {final}})$ against a background of spin orientations created at $T_{\text {initial}}$ after aging for $t_{{\text {w}},{\text {initial}}}$.  To the extent that TC occurs, the chaotic spin arrangement interferes with the growth of $\xi(t_{\text {w,final}})$, reducing its value of $\xi(t_{{\text {w}},{\text {final}}},T_{\text {final}})$ from what it would have been had there been no TC.  If the difference is small, it means that only a small amount of spin orientation after the growth at $T_{\text {final}}$ for $t_{{\text {w}},{\text {final}}}$ is uncorrelated with the spin orientation created at $T_{\text {initial}}$ after $t_{{\text {w}},{\text {initial}}}$.  That small amount must be the amount of a chaotic state created at $T_{\text {final}}$ after the temperature drop.

The relative difference is defined as $\delta_{\text {TC}}$, and exhibited in a percentage form as
\begin{equation}
    \delta_{\text {TC}}=\frac{t_{\text {R}}^{\text {eff}}-t_{{\text {w}},T_{\text {initial}}\rightarrow T_{\text {final}}}^{\text {eff}}}{t_{{\text {w}},T_{\text {initial}}\rightarrow T_{\text {final}}}^{\text {eff}}} \times 100\%.
    \label{eq:delta_TC}
\end{equation}
It is plotted as a function of $T_{\text {final}}$ in Fig. \ref{fig:results}(b).  One sees that, even for the smallest temperature drop in our experiments ($\Delta T = T_{\text {initial}}-T_{\text {final}}=0.05$ K at $T_{\text {final}}=17.95$ K), the calculated characteristic time, $t_{\text {R}}^{\text {eff}}$, is larger than the measured characteristic time, $t_{{\text {w}},T_{\text {initial}}\rightarrow T_{\text {final}}}^{\text {eff}}$, by only around 4\%. For a slightly larger temperature drop, $\Delta T = 0.10$ K, the difference is larger ($\sim$7\%) but still small.  It is clear from the asymptotic slope of Fig. \ref{fig:results}(b) that TC is present, no matter how small $\Delta T$, which is consistent with the conclusions from numerical simulations \cite{Fernandez:2013,Baity:21}.  As seen from Fig. \ref{fig:results}(b), $\delta_{\text {TC}}$ increases rapidly as $\Delta T$ increases ($T_{\text {final}}$ decreases).  This rapid increase in TC exhibited in Fig. \ref{fig:results}(b) occurs very near the temperature at which Ref. \cite{zhai:22} reported the transition to a fully chaotic state, and is consistent with the predictions of the renormalization group \cite{McKay1982,Bray:87}.

The width of $S(t)$ is very broad (examples are exhibited in Appendix A, Figs. \ref{fig:si_s_t_16p5K_frf} and \ref{fig:si_s_t_16p5K_srf}). Hence, even though the characteristic times $t_{\text {R}}^{\text {eff}}$ are much longer that our laboratory measurement time, the ``tail'' of their contribution to $S(t)$ is sufficient to contribute to the experimentally measured characteristic time even for $\Delta T = 0.70$ K ($T_{\text {final}}=17.30$ K).  At that temperature, the measured characteristic time $t_{{\text {w}},T_{\text {initial}}\rightarrow T_{\text {final}}}^{\text {eff}}$ is $13.23\times 10^4$ s while $t_{\text {R}}^{\text {eff}}$ is $38.1\times 10^4$ s, somewhat outside of our laboratory measurement time window. The situation begins to reverse when $t_{\text {R}}^{\text {eff}}$ begins to move rapidly to much larger values as $T_{\text {final}}$ decreases. This is a consequence of both the increase of $\Delta_{\text {max}}$ with decreasing temperature, and the reduction in temperature $T$. For example, the measured characteristic effective time $t_{{\text {w}},T_{\text {initial}}\rightarrow T_{\text {final}}}^{\text {eff}}$, exhibited in Fig. \ref{fig:results}(a), begins to decrease as $\Delta T > 0.70$ K for $T_\text{final} < 17.30$ K.  At $T_\text{final} = 17.00$ K, $t_{\text {R}}^{\text {eff}}$ is two orders of magnitude larger than the native time.  As $\Delta T$ continues to increase, the difference increases rapidly, reaching six orders of magnitude at $\Delta T = 3.00$ K ($T_\text{final}=15.00$ K).
\begin{table*}[htbp]
    \centering
    \caption{Listing of measured and calculated effective response times for $T_{\text {initial}}=16.00$ K at different $T_{\text {final}}$.}
    \begin{ruledtabular}
        \begin{tabular}{cccccccc}
       $T$(K) &$t_{{\text {w}},{\text {native}}}^{\text {eff}}$($\times~10^4$~s) & $t_{{\text {w}},T_{\text {initial}}\rightarrow T_{\text {final}}}^{\text {eff}}$($\times~10^4$ s) & $t_{\text {\text {R}}}^{\text {eff}}$(s) & $\xi/a_0$ \\
        \hline
        16.00  & 1.92 $\pm$ 0.01 & - & - & 7.258 \\
        15.50  & 1.90 $\pm$ 0.03 & 9.70 $\pm$ 0.08 & 1.37~$\times~10^{5}$ & 6.695 \\
        15.00  & 1.89 $\pm$ 0.01 & 12.37 $\pm$ 0.13 & 8.33~$\times~10^{5}$ & 6.183 \\
        14.50  & 1.94 $\pm$ 0.03 & 7.33 $\pm$ 0.11 & 6.45~$\times~10^{6}$ & 5.717 \\
        14.00  & 2.09 $\pm$ 0.01 & 3.36 $\pm$ 0.11 & 5.89~$\times~10^{7}$ & 5.384 \\
        13.00  & 1.84 $\pm$ 0.07 & 1.98 $\pm$ 0.11 & 8.23~$\times~10^{9}$ & 4.625 \\
        12.00  & 1.87 $\pm$ 0.03 & 1.71 $\pm$ 0.04 & 2.62~$\times~10^{12}$ & 4.052 \\
        11.00  & 1.82 $\pm$ 0.01 & 1.49 $\pm$ 0.02 & 2.38~$\times~10^{15}$ & 3.558 \\
        \end{tabular}
    \end{ruledtabular}
    \label{tab:t_1_16K_results}
\end{table*}

The relative difference between $t_{{\text {w}},T_{\text {initial}}\rightarrow T_{\text {final}}}^{\text {eff}}$ and $t_{{\text {w}},{\text{native}}}^{\text {eff}}$ is a measure of how close the system is to a maximally decorrelated state from that prepared at $T_{\text {initial}}$.  We define the relative difference, $\delta_{\mathcal {C}}$, in a percentage form as
\begin{equation}
    \delta_{{\mathcal {C}}}=\frac{t_{{\text {w}},T_{\text {initial}}\rightarrow T_{\text {final}}}^{\text {eff}}-t_{{\text {w}},\text {native}}^{\text {eff}}}{t_{{\text {w}},\text {native}}^{\text {eff}}} \times 100\%.
    \label{eq:delta_C}
\end{equation}
$\delta_{\mathcal {C}}$ is exhibited for the lowest values of $T_{\text {final}}$ in Fig. \ref{fig:results}(c). There are three features that are important.  First, $\delta_{\mathcal {C}}$ is constant for $T_{\text {final}}\leq 14$ K.  This means that the contribution of $t_{\text {R}}^{\text {eff}}$ to $S(t)$ is negligible for $T_{\text {final}}\leqslant 14$ K. The sample is maximally decorrelated from the state prepared at $T_{\text {initial}}$. Second, $\delta_{{\mathcal {C}}}$ is negative.  This is a consequence of the ``imprint'' of the correlations created at $T_{\text {initial}}$ but now completely frozen at $T_{\text {final}}$.  The correlation length $\xi(t_{{\text {w}},{\text {final}}},T_{\text {final}})$ grows from nucleation at $T_{\text {final}}$, but its growth is slowed because of interference with the frozen background correlations created by $\xi(t_{{\text {w}},{\text {initial}}},T_{\text {initial}})$.  This is consistent with the origins of rejuvenation and memory in \cite{Freedberg:23,Paga:23}.  Third, as the temperature is reduced, Eq. (\ref{eq:d_Delta_d_T_exp}) displays an exponential increase in $\Delta_{\text {max}}(t_{\text {w}},T)$. This would result in an exponential reduction of the reversible component of $t_{{\text {w}},{T_{\text {initial}}\rightarrow T_{\text {final}}}}^{\text {eff}}$ as the temperature is lowered, exhibited as the dotted line in Fig. \ref{fig:results}(c).

Complementary experiments were carried out at $T_{\text {initial}}=16.00$ K instead of $T_{\text {initial}}=18.00$ K to check if our analysis is consistent with the accepted length scale dependence of TC.  Table \ref{tab:t_1_16K_results} lists our results for $T_{\text {initial}}=16.00$ K in the same format as in Table \ref{tab:results} for $T_{\text {initial}}=18.00$ K.  First, for $\Delta T = 0.50$ K, the calculated $t_{\text {R}}^{\text {eff}}$ is smaller for $T_{\text {initial}}=16.00$ K ($1.37\times 10^5$ s as compared to $1.70\times 10^5$ s) because, at the lower temperature, the growth of the correlation length is slower, resulting is a smaller $\Delta_{\text {max}}(t_{{\text {w}},{\text {initial}}},T_{\text {initial}})$ for $T_{\text {initial}} = 16.00$ K as compared to $T_{\text {initial}}=18.00$ K.  Concomitantly, $t_{{\text {w}},T_{\text {initial}}\rightarrow T_{\text {final}}}^{\text {eff}}=9.70\times 10^4$ s for $T_{\text {initial}}=16.00$ K is less than that of $10.04\times 10^4$ s for $T_{\text {initial}}=18.00$ K.

When we compare $\delta_{\text {TC}}$ at the same values of $\Delta T$ for the two initial values of $T_{\text {initial}}$, we are comparing the differences in the amount of TC.  We find, expressed as percentages in analogy with Fig. \ref{fig:results}(b), $\delta_{\text {TC}}=41\pm 1\%~(T_{\text {initial}}=16.00$ K) vs $\delta_{\text {TC}} = 69\pm 1\%~(T_{\text {initial}}=18.00$ K), or a lesser amount of TC for $T_{\text {initial}}=16.00$ K by $28\pm 2\%$.  Remembering that TC is related to a comparison between length scales for the correlation length and the equivalent length for chaos \cite{Bray:87}, we can make use of the values for $\xi/a_0$ in the fifth columns of Tables \ref{tab:results} and \ref{tab:t_1_16K_results}.  The difference of $\xi/a_0$ between $T$ = 18.00 K and $T$ = 16.00 K is 9.715 - 7.258 = 2.457, or about 25\% of the former.  Because the chaos length scale exponent $1/\zeta\sim 1.0$ (Appendix B of \cite{zhai:22}) this difference is approximately the difference in the amount of TC between the two initial temperatures.  The consistency of each of these estimates lends further credence to the TC interpretation of our experiments.

\begin{figure}[htbp]
    \centering
    \includegraphics[width=0.48\textwidth]{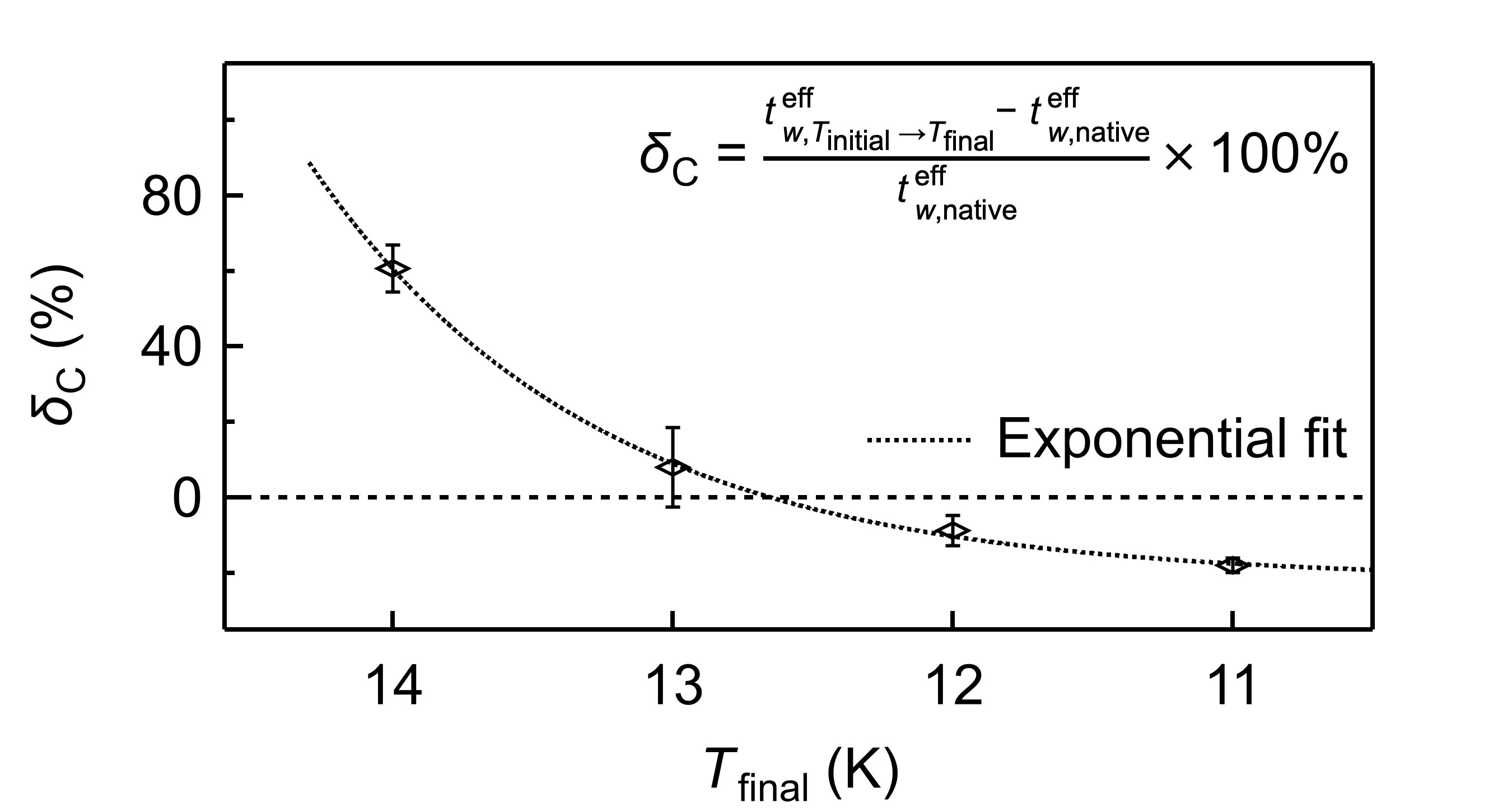}
    \caption{The percent difference $\delta_{{\mathcal {C}}}=(t_{{\text {w}},T_{\text {initial}}\rightarrow T_{\text {final}}}^{\text {eff}}-t_{{\text {w}},\text {native}}^{\text {eff}})/t_{{\text {w}},\text {native}}^{\text {eff}}\times 100\%$ between the measured effective response time after the $T_{\text {initial}} \rightarrow T_{\text {final}}$ protocol and the measured native effective response time at large $\Delta T$s.}
    \label{fig:diff_lDT_16KT1}
\end{figure}

A further comparison of the results for $T_{\text {initial}}=16.00$ K with $T_{\text {initial}}=18.00$ K, but now at larger $\Delta T$, illustrates the approach to the compound state approaching maximally decorrelization with the initially prepared state through $\delta_{{\mathcal {C}}}$. Expressed as a percentage, Table \ref{tab:percent_differences} compares $\delta_{\mathcal {C}}$ for the two values of $T_{\text {initial}}$. The difference $\delta_{\mathcal {C}}$ for $T_{\text {initial}}=16.00$ K is plotted in Fig. \ref{fig:diff_lDT_16KT1}.  As before, the approach to complete decorrelization can be fitted to an exponential (the dotted curve in Fig.  \ref{fig:diff_lDT_16KT1}). As $\Delta T$ increases from $T_{\text {initial}}=16.00$ K, the approach to a completely decorrelated state is slower than that at $T_{\text {initial}}=18.00$ K.  This is because $\xi/a_0 (T=16.00$ K) is smaller than $\xi/a_0 (T=18.00$ K) (see Tables \ref{tab:results} and \ref{tab:t_1_16K_results}), requiring a larger $\Delta T$ for crossover to chaos \cite{Bray:87}.  This is seen quantitatively in Table \ref{tab:percent_differences} where it takes a larger $\Delta T$ for $t_{{\text {w}},T_{\text {initial}}\rightarrow T_{\text {final}}}^{\text {eff}}$ to reach $t_{{\text {w}},{\text {native}}}^{\text {eff}}$ for $T_{\text {initial}}=16.00$ K as compared to that for $T_{\text {initial}}=18.00$ K.
\begin{table}[htbp]
    \centering
    \caption{Listing of the percent difference $\delta_{{\mathcal {C}}}=(t_{{\text {w}},T_{\text {initial}}\rightarrow T_{\text {final}}}^{\text {eff}}-t_{{\text {w}},\text {native}}^{\text {eff}})/t_{{\text {w}},\text {native}}^{\text {eff}}\times 100\%$ at different $\Delta T$s between $T_{\text {initial}}$ = 18.00 K and $T_{\text {initial}}$ = 16.00 K.}
    \begin{ruledtabular}
        \begin{tabular}{cccccccc}
       $\Delta T$(K) & $T_{\text{initial}}$(K) & $T_{\text{final}}$(K) & $\delta_{\mathcal {C}}$(\%) \\
        \hline
       \multirow{2}{*}{1.00}  & 18.00 & 17.00 & 480 $\pm$ 23 \\
         & 16.00 & 15.00 & 554 $\pm$ 8 \\
         \hline
         \multirow{2}{*}{1.50}  & 18.00 & 16.50 & 155 $\pm$ 10 \\
         & 16.00 & 14.50 & 278 $\pm$ 11 \\
         \hline
         \multirow{2}{*}{2.00}  & 18.00 & 16.00 & 45 $\pm$ 3 \\
         & 16.00 & 14.00 & 61 $\pm$ 6 \\
         \hline
         \multirow{2}{*}{3.00}  & 18.00 & 15.00 & -8 $\pm$ 1 \\
         & 16.00 & 13.00 & 8 $\pm$ 11 \\
         \hline
         \multirow{2}{*}{4.00}  & 18.00 & 14.00 & -15 $\pm$ 2 \\
         & 16.00 & 12.00 & -9 $\pm$ 4 \\
         \hline
        \multirow{2}{*}{5.00}  & 18.00 & 13.00 & -17 $\pm$ 7 \\
         & 16.00 & 11.00 & -18 $\pm$ 2 \\
        \end{tabular}
    \end{ruledtabular}
    \label{tab:percent_differences}
\end{table}

\section{\label{sec:summary}Summary}
Our experiments display the growth of TC in spin glasses over the full range of temperature change.  We posit that a reduction in temperature results in a compound state composed of a reversible component, and a component that is maximally decorrelated from the state originally created at $T_{\text {initial}}(t_{\text {w}}, T)$.  The magnitude of the latter is small at small $\Delta T$.  As the reduction in temperature increases, the latter becomes more significant, increasing rapidly in the vicinity of the crossover $\Delta T$ predicted from renormalization group methods \cite{Bray:87} and seen experimentally \cite{Guchhait:15,zhai:22}.  Upon further increase in $\Delta T$, the reversible component rapidly decreases, leaving the maximally decorrelated component dominant. We believe this set of experiments lays a firm basis for the onset of TC in spin glass dynamics, and confirms the predictions of simulations that find ``... a complete reorganization of the equilibrium configurations [takes place] by the slightest change in temperature.''  

\section*{Acknowledgments}
We are pleased to acknowledge the assistance of the Janus II Collaboration during the progress of our experiments, and the advice in particular of Professor Victor Martin-Mayor.  We also acknowledge the suggestions for our data analysis from Professor E. Dan Dahlberg and Dr. J. Freedberg.  The use of the single CuMn crystal, grown by Dr. D.L. Schlagel at Ames Laboratory, was crucial for our investigation.  This work was supported by the U.S. Department of Energy, Office of Science, Basic Energy Sciences, Division of Materials Science and Engineering, under Award No. DE-SC0013599.  Dr. D.L. Schlagel's work was performed at Ames Laboratory, which is operated for the U.S. Department of Energy by Iowa State University under Contract No. DE-AC02-07CH11358. H. Li acknowledges the partial support by the National Science Foundation through the Center for Dynamics and Control of Materials: an NSF MRSEC under Cooperative Agreement No. DMR-2308817.

\appendix
\section{Determination of the $S(t)$ peak}
As shown in Eq. (\ref{eq:s_of_t}),  the relaxation function $S(t)$ is the first derivative of the magnetization with respect to the natural logarithm of time. Therefore, the noise in the magnetization measurements makes it impractical to locate the $S(t)$ peak by using raw data. In our analysis, we used a log-normal distribution function to obtain a smooth approximation to the raw data.
\begin{equation}
M_{\text {ZFC}}(t,t_{\text {w}};T)=M_0+Ae^{-[\text {ln}(t/t_0)/w]^2}
\label{eq:si_log_normal}
\end{equation}

\begin{figure}[htbp]
    \centering
    \includegraphics[width=7.75cm]{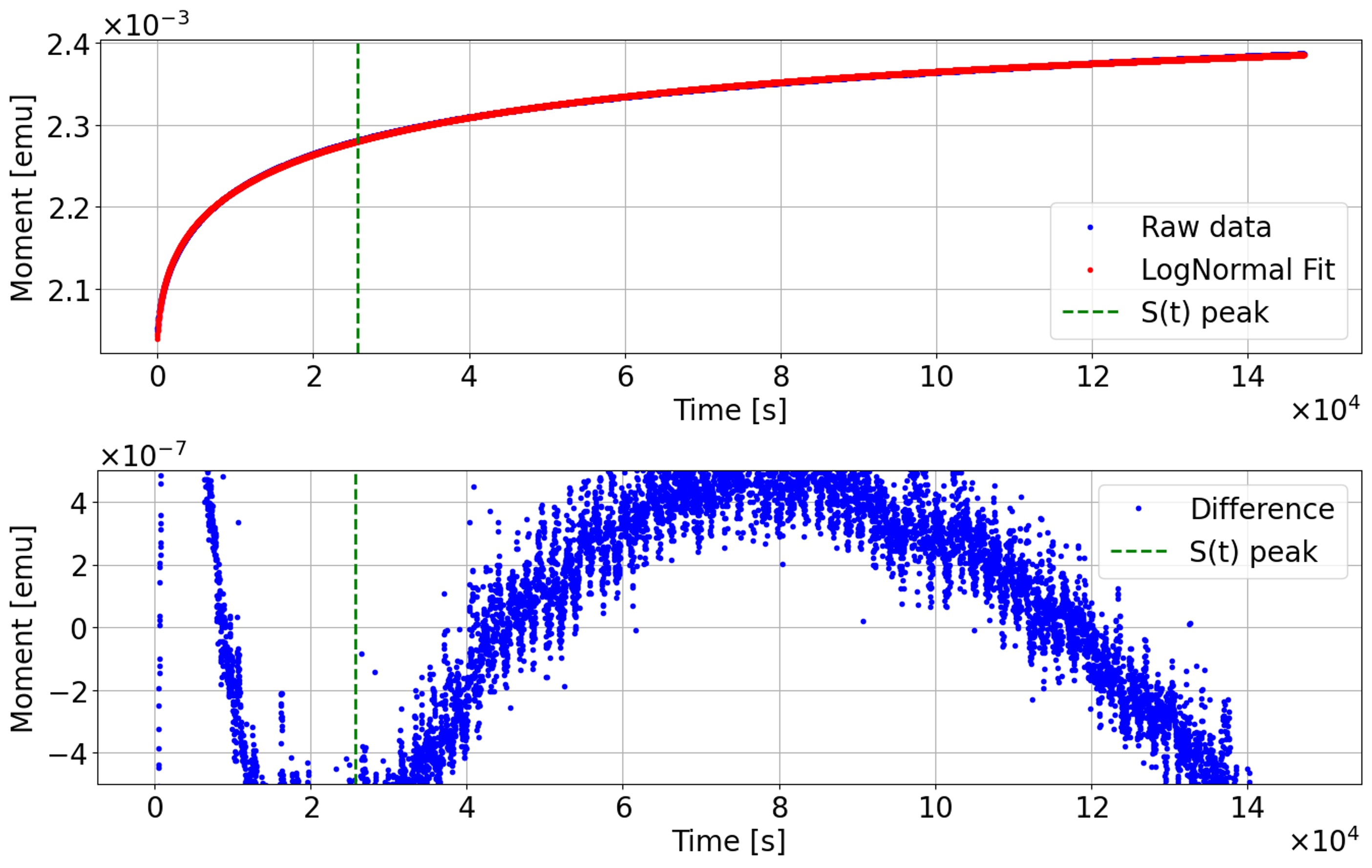}
    \caption{(Color online) The full-range fit of the native measurement data at $T_{\text {final}} = 16.50$ K to Eq. (\ref{eq:si_log_normal}). The upper panel shows the fit result and the lower panel shows the difference between the fit result and the raw data.  The extracted peak position is 2.57~$\times~10^4$ s as indicated by the green dash line, where the difference between the fit and raw data is largest.}
    \label{fig:si_16p5K_frf}
\end{figure}

\begin{figure}[htbp]
    \centering
    \includegraphics[width=7.5cm]{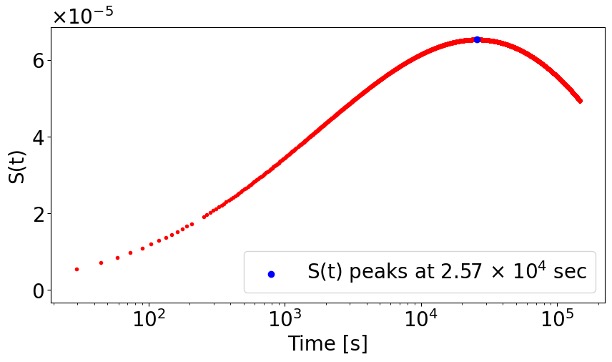}
    \caption{(Color online) $S(t)$ from the full-range fit of the native measurement at $T_{\text {final}} = 16.50$ K.}
    \label{fig:si_s_t_16p5K_frf}
\end{figure}

The approximation from Eq. (\ref{eq:si_log_normal}) is unable to reproduce the shape of the $M-t$ curve in its full range. Taking the native measurement at $T_{\text {final}} = 16.50$ K as an example, the full-range fit of the raw data to Eq. (\ref{eq:si_log_normal}) in Fig. \ref{fig:si_16p5K_frf} gives an $S(t)$ peak at 2.57~$\times~10^4$ s in Fig. \ref{fig:si_s_t_16p5K_frf}. However, the difference between the fit result and the raw data in the vicinity of the $S(t)$ peak is so large that the position of the $S(t)$ maximum is not reliable.

\begin{figure}[htbp]
    \centering
    \includegraphics[width=7.75cm]{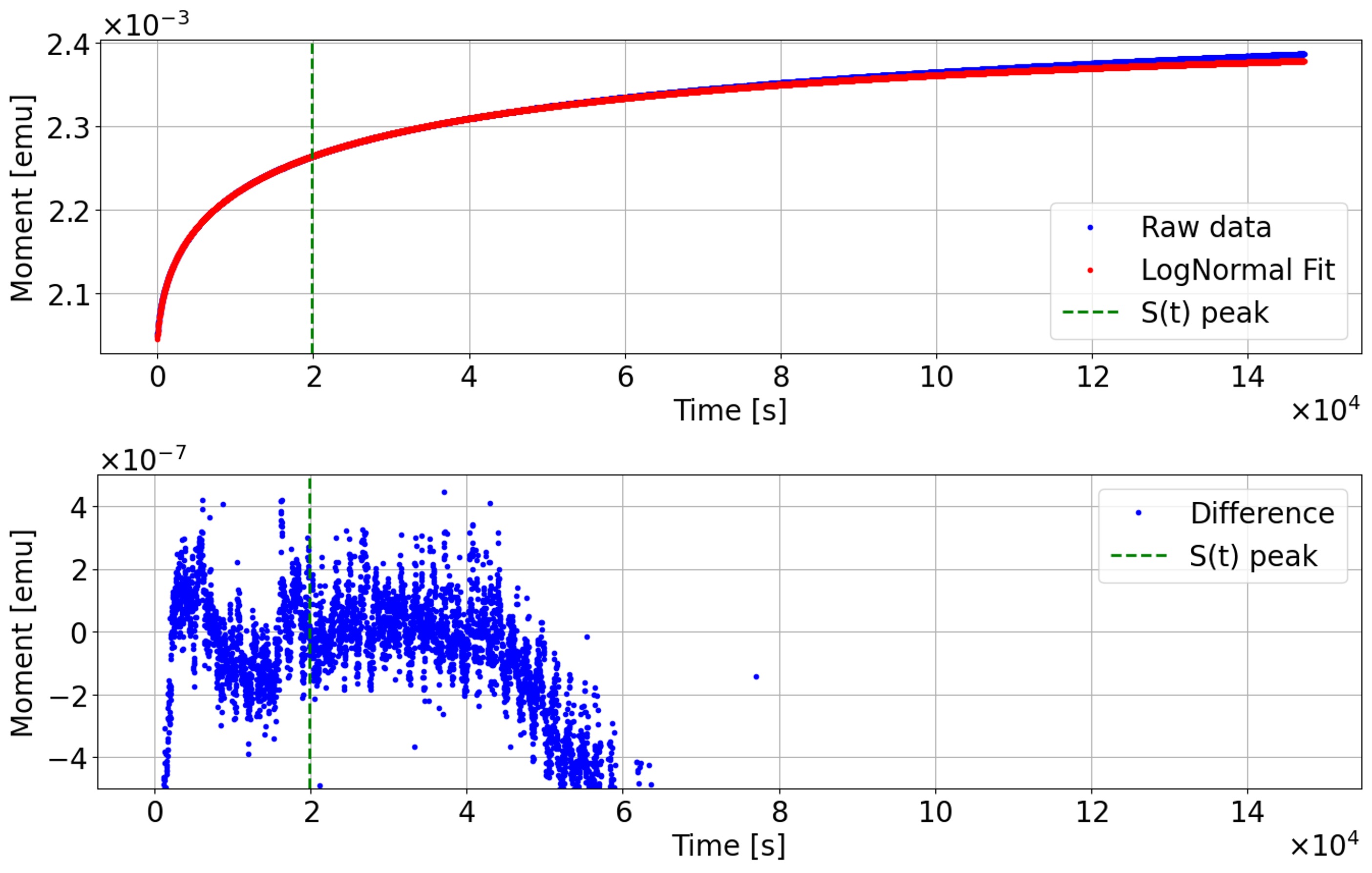}
    \caption{(Color online) The shorter-range fit of the native measurement data at $T_{\text {final}} = 16.50$ K to Eq. (\ref{eq:si_log_normal}). The upper panel shows the fit result and the lower panel shows the difference between the fit result and the raw data. The peak position is at 1.99~$\times~10^4$ s as indicated by the green dash line, where the difference between the fit and raw data is smallest.}
    \label{fig:si_16p5K_srf}
\end{figure}
\begin{figure}
    \centering
    \includegraphics[width=7.5cm]{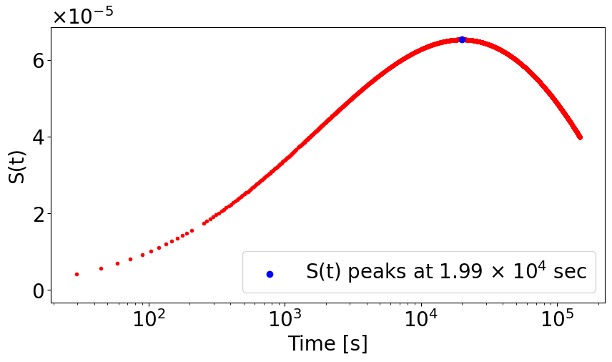}
    \caption{(Color online) $S(t)$ from the short-range fit of the native measurement at $T_{\text {final}} = 16.50$ K.}
    \label{fig:si_s_t_16p5K_srf}
\end{figure}

Considering that we are only interested in the position of the maximum of the $S(t)$ curve, a shorter range fit around the expected $S(t)$ peak in Fig. \ref{fig:si_16p5K_srf} is a more attractive approach. For example, we find the $S(t)$ peak position from a shorter-range fit to be 1.99~$\times~10^4$ as shown in Fig. \ref{fig:si_s_t_16p5K_srf}. The difference between the fit result and the raw data near $t = 1.99\times 10^4$ s indicates a good fit and thus a reliable $S(t)$ peak position.

In the short-range fit, the size and position of the fitting range could affect the position of the $S(t)$ peak. For that reason, we take the center of the shorter-range fit to be at the expected position of the $S(t)$ peak.  The fitting range should be large enough that it covers the earlier period where $M_{\text {ZFC}}(t,t_{\text {w}};T)$ increases relatively rapidly, so that, as much as possible, the shape information of the $M-t$ curve is preserved.  We utilize a scanning protocol to determine the fitting range position.  The $T_{\text {initial}}\rightarrow T_{\text {final}}$ measurement at $T_{\text {final}} = 17.50$ K, as described below, is an example.

\begin{figure}[htbp]
    \centering
    \includegraphics[width=7.5cm]{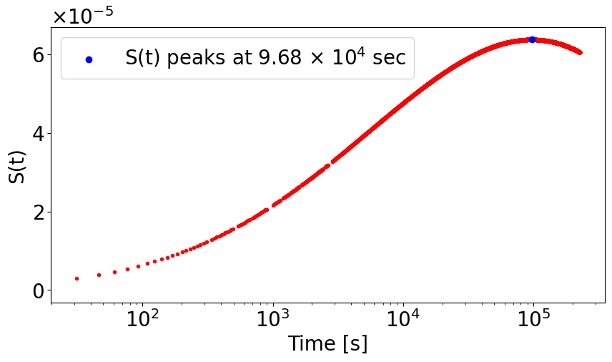}
    \caption{(Color online) $S(t)$ from the full-range fit of the $T_{\text {initial}}\rightarrow T_{\text {final}}$ measurement at $T_{\text {final}} = 17.50$ K.}
    \label{fig:si_s_t_17p5K_frf}
\end{figure}

(1) We first fit all points in our raw data to Eq. (\ref{eq:si_log_normal}) to estimate the position of the $S(t)$ peak. At $T_{\text {final}} = 17.50$ K, the full-range fit gives an $S(t)$ peak at 9.68~$\times~10^4$ s as shown in Fig. \ref{fig:si_s_t_17p5K_frf}, which is the 5437$^\text{th}$ point in our raw data.
\begin{table*}[htbp]
    \centering
    \caption{Listing of measured native effective response time and $T_{\text {initial}}\rightarrow T_{\text {final}}$ effective response time for the temperature $T_{\text {final}}$ = 17.80 K at different applied magnetic fields.}
    \begin{ruledtabular}
        \begin{tabular}{cccccccc}
       $T_{\text {initial}}$(K) & $T_{\text {final}}$(K) & $H$(Oe) &$t_{\text {w},{\text {native}}}^{\text {eff}}$($\times~10^4$~s) & $t_{\text {w},T_{\text {initial}}\rightarrow T_{\text {final}}}^{\text {eff}}$($\times~10^4$ s)\\
        \hline
        \multirow{2}{*}{18.00} & \multirow{2}{*}{17.80} & 50  & 1.96 $\pm$ 0.01 & 5.42 $\pm$ 0.08  \\
          &   & 100 & 1.90 $\pm$ 0.03 & 5.42 $\pm$ 0.10 \\
        \end{tabular}
    \end{ruledtabular}
    \label{tab:50Oe_results}
\end{table*}
\raggedbottom

\begin{figure}[htbp]
    \centering
    \includegraphics[width=7.75cm]{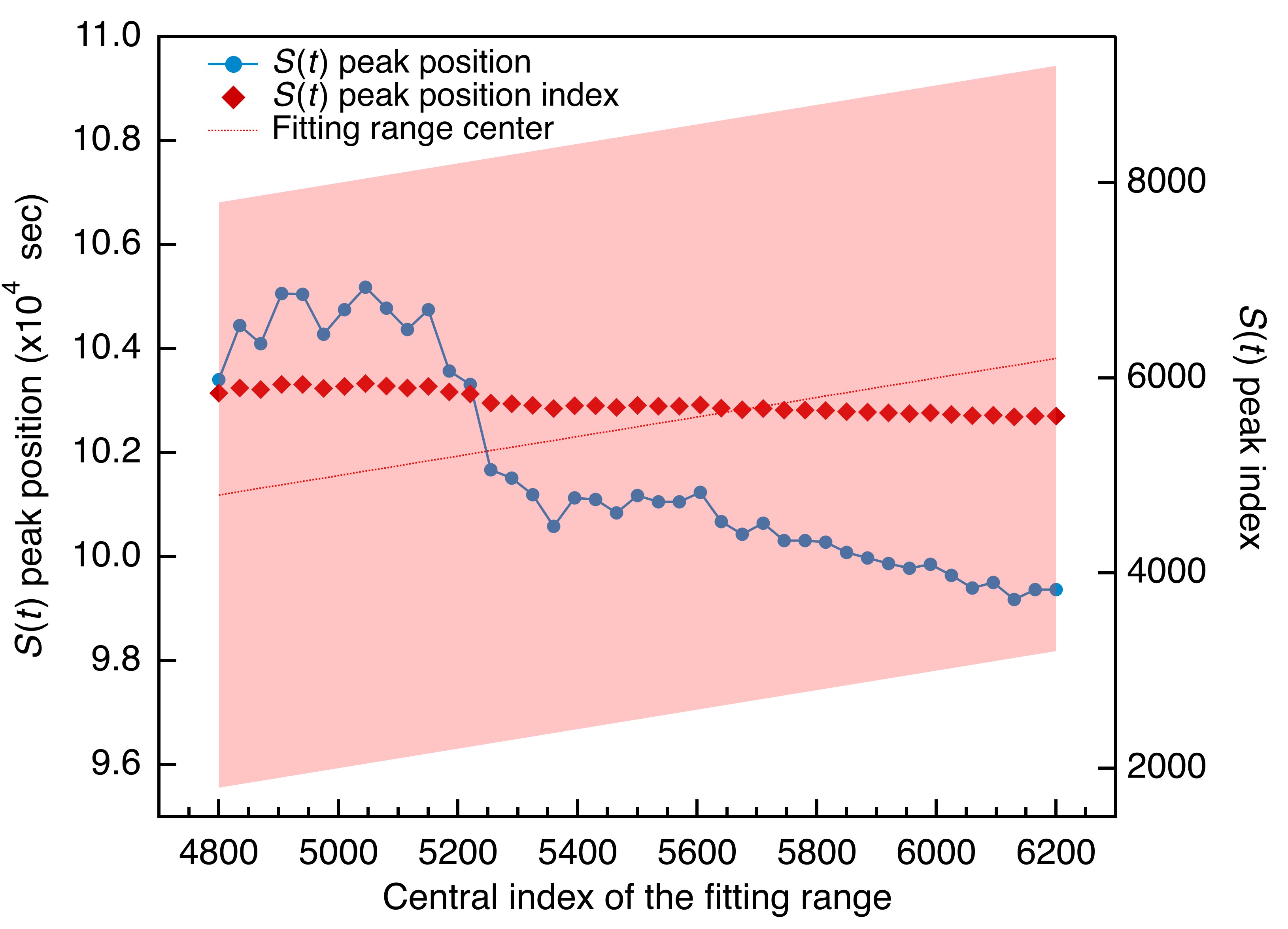}
    \caption{(Color Online) The short-range fit scanning results of the $T_{\text {initial}}\rightarrow T_{\text {final}}$ measurement at $T_{\text {final}} = 17.50$ K. The blue dots show the $S(t)$ position and the red diamonds show their corresponding index in the raw data. The red shaded region shows the fitting range, with the center of the fitting range depicted by the red dashed line.}
    \label{fig:si_fitting_scan}
\end{figure}

(2) We then choose the size of the fitting range for the short-range fit. In our measurements, the time interval between each data point is approximately 15 s. Typically a 2001-point fitting range is large enough for our native measurement, while the $T_{\text {initial}}\rightarrow T_{\text {final}}$ measurements require a 2001-point to 6001-point fitting range depending on the $S(t)$ peak position. In our analysis, we found that a reasonable change (5\%) in the size of the fitting range still provides us with similar results for the $S(t)$ peak position. The difference between the results from the original fitting range and the adjusted fitting range is taken as the error bars. For the $T_{\text {initial}}\rightarrow T_{\text {final}}$ measurement at $T_{\text {final}} = 17.50$ K, we used a 6001-point fitting range. We scanned the 6001-point fitting range across our raw data from centering at the 4800$^\text{th}$ point to centering at the 6200$^\text{th}$ point with a 35-point interval. The $S(t)$ peak position and its data point index are shown in Fig. \ref{fig:si_fitting_scan}. Comparing the center of the fitting range and the index of the corresponding $S(t)$ peak position, we found that the center of the fitting range should be near the 5670$^\text{th}$ point.


\begin{figure}[htbp]
    \centering
    \includegraphics[width=7.5cm]{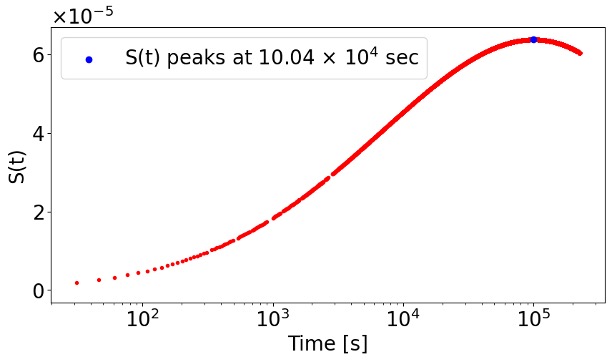}
    \caption{(Color Online) $S(t)$ from the 6001-point fit of the $T_{\text {initial}}\rightarrow T_{\text {final}}$ measurement at $T_{\text {final}} = 17.50$ K, with the $S(t)$ peak position being the center of the fitting range.}
    \label{fig:si_s_t_17p5K_srf}
\end{figure}

(3) We take an iterative process to determine the final position of the fitting range. As discussed in (2), the center of the fitting range should be near the 5670$^\text{th}$ point. Therefore, we started the following iterative process: we started with a fitting range from the 2670$^\text{th}$ point to the 8670$^\text{th}$ point, whose center is the 5670$^\text{th}$ point, and obtained a $S(t)$ peak of 10.04~$\times~10^4$ s at the 5674$^\text{th}$ point in our raw data; we then set the 5674$^\text{th}$ point as the center of our new fitting range from the 2674$^\text{th}$ point to the 8674$^\text{th}$ point, and obtained a $S(t)$ peak of 10.04~$\times~10^4$ s at the 5674$^\text{th}$ point in our raw data. With the fitting range center being exactly at the $S(t)$ peak position, we consider it our final result of the $S(t)$ peak position, as exhibited in Fig. \ref{fig:si_s_t_17p5K_srf}. 

Using a shorter-range fit of the raw data to Eq. (\ref{eq:si_log_normal}) with the scanning protocol explained above, we are able to obtain the characteristic time from the $S(t)$ peak position at different temperatures with much higher accuracy than has heretofore been reported in the literature.  The entries in Table \ref{tab:results} and Table \ref{tab:t_1_16K_results} of the main text are the results of this process.

\section{The magnetic linearity}
To investigate the magnetic linearity of our results, we performed measurements for the native effective response time, $t_{\text {w},{\text{native}}}^{\text {eff}}$, and the $T_{\text {initial}}\rightarrow T_{\text {final}}$ effective response time, $t_{\text {w},T_{\text {initial}}\rightarrow T_{\text {final}}}^{\text {eff}}$, with $T_{\text {initial}}$ = 18.00 K and $T_{\text {final}}$ = 17.80 K under an applied magnetic field of 50 Oe. The results obtained at 50 Oe are listed in Table \ref{tab:50Oe_results} with the results obtained at 100 Oe, and the magnetic field effect on both effective response times is negligible in our analysis and discussions.

\bibliography{citations}

\end{document}